\documentclass[twocolumn,showpacs,amsmath,amssymb]{revtex4}
\usepackage[dvips]{graphics}
\def\imo{i}
\def\re#1{Re(#1)}
\def\im#1{Im(#1)}

\newcommand{\arXivonly}[1]{#1}

\begin{document}
\title{Superradiance and instability of the charged Myers-Perry black holes\\ in the G\"{o}del Universe}
\author{R. A. Konoplya}\email{konoplya_roma@yahoo.com}
\affiliation{DAMTP, Centre for Mathematical Sciences, Wilberforce Road, Cambridge CB3 0WA, UK.}
\author{A. Zhidenko}\email{olexandr.zhydenko@ufabc.edu.br}
\affiliation{Centro de Matem\'atica, Computa\c{c}\~ao e Cogni\c{c}\~ao, Universidade Federal do ABC,\\ Rua Santa Ad\'elia, 166, 09210-170, Santo Andr\'e, SP, Brazil}
\affiliation{Theoretical Astrophysics, Eberhard-Karls University of T\"{u}bingen, T\"{u}bingen 72076, Germany}
\begin{abstract}
We consider scalar field perturbations of the asymptotically G\"odel 5-dimensional charged rotating black holes with two equal angular momenta.
It is shown that the spectrum of proper oscillations of the perturbation includes superradiant unstable modes. The reason for the instability is the
confining Dirichlet boundary condition at the asymptotically far region of the G\"odel Universe. The confining box makes superradiant modes extract rotational energy from the black hole and, after repeated reflections from the black hole, grow unboundedly. A similar instability takes place for rotating black holes in the asymptotically anti-de Sitter (AdS) space-time.
\end{abstract}
\maketitle

\section{Introduction}

Stability of a black hole's space-time against small perturbations is the basic requirement for its existence. Stability acquires
a special meaning in higher dimensional gravity, because there is no traditional uniqueness theorem for higher dimensional solutions with an event horizon. A number of ``black'' solutions with various topologies have been found \cite{Emparan:2008eg} and stability could be the criterium which could discard unphysical solutions \cite{Ishibashi:2011ws,Konoplya:2011qq}. Through the gauge/gravity duality, classical instability of black holes can be interpreted as the onset of thermodynamical phase transition in the dual field theory in some cases \cite{Gubser:2000mm}.
Therefore, stability of various black holes has been actively studied during past decade \cite{stability}.
In most cases it is difficult to prove (in)stability of black holes analytically. Then, analysis of its proper oscillations, described by the called quasinormal modes, is used: If no unstable modes are found in the spectrum, the space-time is believed to be stable.

Stability of various higher dimensional black holes have been proved in  \cite{stability,instability} in the context of string theory and higher dimensional gravity. Instability, once it happens, can be stipulated by different reasons. Various instabilities have been observed for higher dimensional black holes in \cite{instability}. A special kind of instability takes place for rotating black holes in the AdS space-time, because of the phenomenon called the superradiance \cite{Starobinsky}.
The latter is the amplification of the reflected wave due to extraction of rotational energy from the black hole. If the perturbation propagate under the Dirichlet boundary condition at infinity, what occurs for asymptotically AdS black holes, the superradiance naturally leads to instability.
Superradiance and the instability induced by it have been studied in a great number of works (see for instance \cite{Konoplya:2011qq,superradiance} and references therein). It was shown \cite{Kodama:2009rq} that all (and only) superradiant modes are unstable for gravitational perturbation of the higher dimensional simply rotating asymptotically AdS black hole.

In the previous paper \cite{Konoplya:2011ig} we considered the proper oscillation frequencies, called quasinormal modes, of the non-rotating 5-dimensional black hole immersed in the higher dimensional analogue of the G\"odel (rotating) Universe \cite{Godel,Gimon:2003ms}. The exact solution for the black hole metric was found by Gimon and Hashimoto in the supergravity \cite{Gimon:2003ms} and was further investigated in a number of works \cite{string-pp}.
In \cite{Konoplya:2011ig} we have shown that the quasinormal spectrum is similar to the spectrum of normal modes of the pure G\"odel space-time.
The latter has a number of common features with the spectrum of pure AdS space-time, where the scale of the universe's rotation $j$  plays
the role of the inverse anti-de Sitter radius.

In the present research we shall consider quasinormal modes of a much more general space-time: 5-dimensional charged rotating black holes with both equal angular momenta in the G\"odel Universe. Our main result here is observation of the superradiant instability for such black holes. In comparison with the earlier work \cite{Konoplya:2011ig}, we have two extra parameters, the black hole's rotational parameters $a$ and the black hole charge $Q$. We shall analyse the dependence of the quasinormal modes, and first of all of unstable superradiant modes, from these parameters and determine the parametric regions of instability.

The paper is organized as follows: Sec II gives basic formulas for the 5-dimensional charged rotating black hole in the G\"odel Universe. Sec III is devoted to separation of variables in the test scalar field equation, Sec IV briefly describes numerical methods which we used for analysis of quasinormal modes. Finally, in Sec V we discuss the obtained results and parametric regions of instability.

\section{Charged rotating asymptotically G\"odel black holes}

The bosonic fields of the minimal (4+1)- supergravity theory consist of the metric and the one-form gauge field, which are governed by the following equations of motion
\begin{eqnarray}\label{EinsteinEq}
R_{\mu \nu} &=& 2 \left(F_{\mu \alpha} F_{\nu}^{\alpha} -\frac{1} {6} g_{\mu \nu}
F^{2}\right)\,;\\\label{MaxwellEq}
D_{\mu} F^{\mu \nu} &=& \frac{1}{2 \sqrt{3}} \varepsilon^{\alpha \lambda
\gamma \mu \nu}  F_{\alpha \lambda} F_{\gamma \mu}\,;
\end{eqnarray}
here, $\varepsilon_{\alpha \lambda \gamma \mu \nu} = \sqrt{-\det g_{\mu\nu}}~\epsilon_{\alpha \lambda \gamma \mu \nu}$.

In the Euler coordinates $(t, r, \theta, \psi, \phi)$, the solution for the 5-dimensional charged rotating asymptotically G\"odel black hole with two equal angular momenta $a = a_1 =a_2$ is given by \cite{Wu:2007gg}
\begin{eqnarray}
ds^2 &=& - f(r) dt^2 -q(r) r \sigma_{L}^{3} d t - h(r) r^2
(\sigma_{L}^{3})^2 + \frac{d r^2}{v(r)} \nonumber\\
&&+\frac{r^2}{4}(d \theta^{2} + d \psi^{2} + d \phi^{2} + 2 \cos \theta d \psi d \phi),\label{metric}
\end{eqnarray}
where $\sigma_{L}^{3}= d \phi + cos \theta d \psi$,
\begin{eqnarray}\nonumber
q(r) &=& 2 j r+\frac{6 j Q}{r}+\frac{a(2 M - Q)}{r^3}-\frac{aQ^2}{r^5},\\\nonumber
h(r)&=&j^2 (r^2\! +\! 2 M\! +\! 6Q)\!-\!\frac{3jQa}{r^2}\!-\!\frac{a^2(M\!-\!Q)}{2r^4}\!+\!\frac{a^2Q^2}{4r^6},\\
f(r)&=& 1- \frac{2 M}{r^2}+\frac{Q^2}{r^4},\label{coeff}\\\nonumber
v(r) &=& 1 - \frac{2 M}{r^2} + \frac{8 j (M+Q)(a+2jM+4jQ)}{r^2}\\\nonumber&+&\frac{2(M-Q)a^2 + Q^2 (1 - 16 j a - 8 j^2 M -24j^2 Q)}{r^4}.
\end{eqnarray}

Here $M$ and $Q$ are charge and mass of the black hole. When $Q= 0$ and $a=0$, the above solution is reduced to the Gimon-Hashimoto solution. When $a=j=0$ we have the 5-dimensional Reissner-Nordstr\"om solution. For $Q=0$ and $j=0$ we obtain the Myers-Perry black hole with two equal angular momenta.

\section{Separation of variables}

In order to derive the wave equation one can use the relation
$$q^2(r)+f(r)(1-4h(r))=v(r)\,,$$
which implies
$$g=-\frac{r^6\sin^2\theta}{64}\,.$$

Perturbations of the scalar field  in a curved background are governed by
the Klein-Gordon equation
\begin{equation}\label{Klein-Gordon}
\Box \Phi  \equiv \frac{1}{\sqrt{-g}} \partial_\mu\left(g^{ \mu \nu} \sqrt{-g}
\partial_\nu\Phi\right)  = \mu^2\Phi.
\end{equation}
Since the background metric has the Killing vectors  $\partial_{t}$,
$\partial_{\psi}$,  $\partial_{\phi}$, one can choose the ansatz for the wave function as
\begin{equation}\label{ansatz}
\Phi(t,r,\theta,\psi,\phi) = e^{\displaystyle-\imo \omega t + \imo n \psi + \imo m \phi} Y(\theta) R(r)r^{-3/2}.
\end{equation}

Substituting (\ref{ansatz}) into (\ref{Klein-Gordon}) and separating the variables, one can find that the angular part of the function satisfies the equation
\begin{equation}\label{angular}
\left(\frac{1}{\sin\theta}\frac{d}{d\theta}\sin\theta\frac{d}{d\theta}+\frac{2mn\cos\theta-m^2-n^2}{\sin^2\theta}+\lambda\right)Y(\theta)=0,
\end{equation}
where $\lambda$ is the separation constant with the eigenvalues
$$\lambda=\ell(\ell+1), \quad \ell=\max(|m|,|n|)+i, \quad i=0,1,2\ldots.$$

Then, the equation for the radial part takes the wave-like form
\begin{equation}\label{wave-like}
\left(\frac{d^2}{dr_\star^2}+Q(r_\star)\right)R(r_\star)=0,
\end{equation}
where $r_\star$ is the tortoise coordinate, which is defined as
\begin{equation}\label{tortoise}
dr_\star=\frac{dr}{v(r)}.
\end{equation}
The effective potential can be written in terms of the coordinate $r$ as follows \cite{Konoplya:2011ig}
\begin{eqnarray}
&&Q(r)=\left(1-4h(r)\right)\left(\omega-\frac{2mq(r)}{r(1-4h(r))}\right)^2-\\\nonumber
&&v(r)\left(\frac{4\lambda}{r^2}+\mu^2+\frac{16m^2h(r)}{(1-4h(r))r^2}+\frac{3v(r)+6r v'(r)}{4r^2}\right)\,.
\end{eqnarray}

Now we are in position to perform analysis of the quasinormal spectrum for the obtained wave equation.

\begin{figure*}
\resizebox{\linewidth}{!}{\includegraphics*{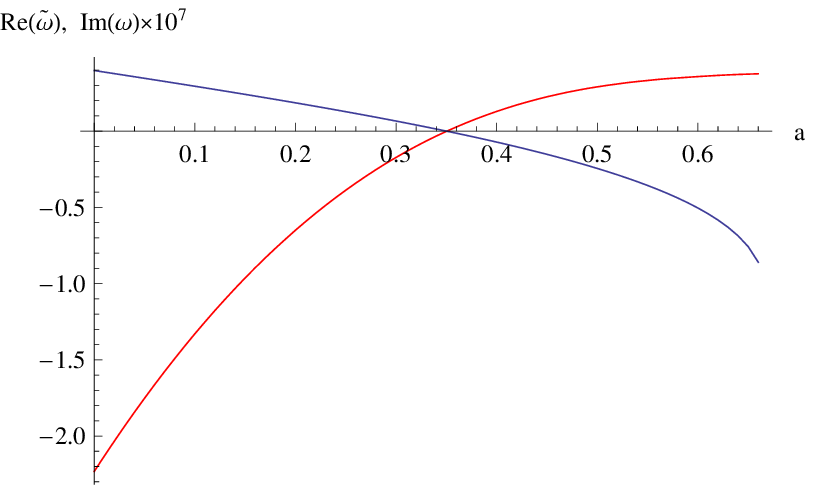}\includegraphics*{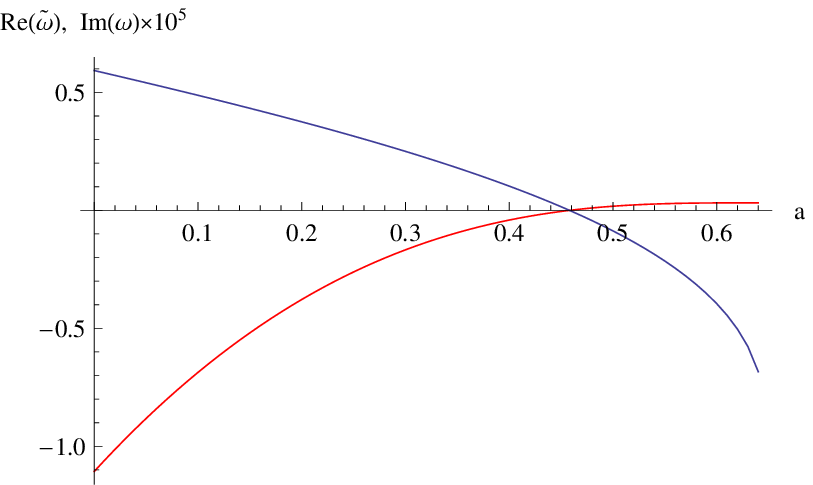}\includegraphics*{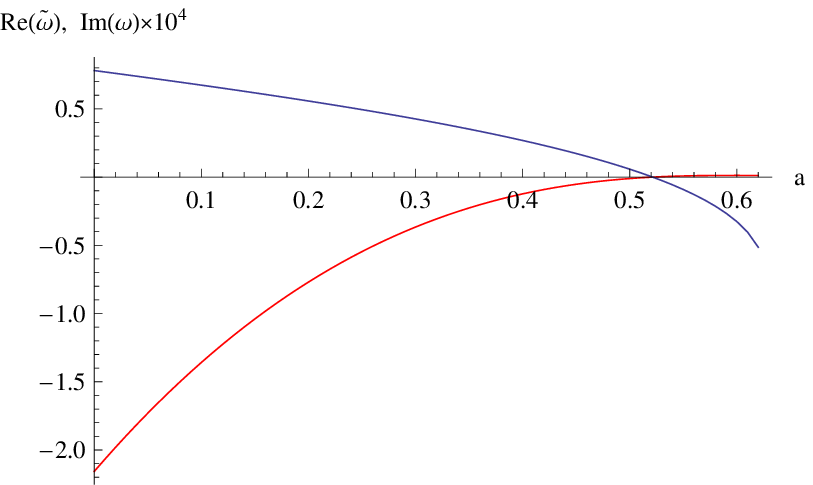}}
\caption{Real (blue) and imaginary (red) parts of $\tilde{\omega}$ as functions of $a$ for $m=1$, $\lambda=2$ ($M=1$, $Q=0$), $j=0.02$ (left panel), $j=0.03$ (middle panel), $j=0.04$ (right panel).}\label{fig:instex}
\end{figure*}

\section{Boundary conditions and numerical methods}

There are two positive solutions of the equation $v(r)=0$: the event horizon $r_+$ and the inner horizon $r_-$
\begin{eqnarray}
2r_\pm^2&=&2 M - 8 a j M - 16 j^2 M^2 - 8 a j Q\\\nonumber
&-&48 j^2 M Q - 32 j^2 Q^2 \pm \sqrt{P}\qquad (r_+>r_->0),
\end{eqnarray}
where
$$P=(2 M - 8 a j M - 16 j^2 M^2 - 8 a j Q - 48 j^2 M Q - 32 j^2 Q^2)^2$$
$$- 4 (2 a^2 M - 2 a^2 Q + Q^2 - 16 a j Q^2 - 8 j^2 M Q^2 - 24 j^2 Q^3).$$
At the classical level, a wave cannot be emitted from the event horizon of a black hole, so that the standard boundary condition for a great variety of black holes is requirement of the purely ingoing wave at the event horizon,
$$ R\propto e^{-\imo\tilde{\omega} r_\star\sqrt{1-4h(r_\star)}}, \quad r_\star\rightarrow-\infty,$$
\begin{equation}\label{hor-ingoing}
 \quad \tilde{\omega}=\omega-\frac{2mq(r_+)}{r_+-4r_+h(r_+)}=\omega-m\Omega_+.
\end{equation}

With respect to the coordinate $r$ (\ref{hor-ingoing}) reads
\begin{equation}\label{hor-asymptot}
R\propto \left(r-r_+\right)^{-\imo\tilde{\omega}b}, \quad r\rightarrow r_+,\quad b=\frac{\sqrt{1-4h(r_+)}}{v'(r_+)}.
\end{equation}

At spatial infinity we have
\begin{equation}\label{inf-asymptot}
R(r\rightarrow\infty)=C_+\Psi_+(r)+C_-\Psi_-(r),
\end{equation}
where
$$\Psi_\pm(r)=e^{\pm j\omega r^2}r^{\alpha_\pm}\left(1+\frac{A_{1\pm}}{r}+\frac{A_{2\pm}}{r^2}+\frac{A_{3\pm}}{r^3}\ldots\right),$$
with
\begin{eqnarray}
\alpha_\pm&=&-\frac{1}{2}\pm K,\\
K&=&2m- \frac{\omega^2-\mu^2}{4\omega j}\\\nonumber
&+&2j\omega(M+Q)\left(3- 8ja - 16(M+2Q)j^2\right).
\end{eqnarray}

Since the exponents $e^{\pm j\omega r^2}$ have purely real index, they do not describe in-going or out-going waves. Therefore, we are unable to impose usual quasi-normal boundary conditions. However, we can use the analogy with AdS backgrounds and require Dirichlet boundary conditions at spatial infinity. This implies that $C_+=0$ for $\re{j\omega}>0$ or $C_-=0$ for $\re{j\omega}<0$.

The non-trivial behavior of the functions $\Psi_\pm$ is observed when $\re{\omega}=0$. In this case both exponents have oscillatory behavior at spatial infinity. Thus, in order to impose the Dirichlet boundary conditions we must consider the factor $r^{\alpha_\pm}$. When $\re{\omega}=0$ one can find that $$\re{\alpha_\pm}=-\frac{1}{2}\pm2m,$$
implying that one of $\Psi_\pm$ is convergent and the other one is divergent as $r\rightarrow\infty$. The only exception is $m=0$, when both $\Psi_+$ and $\Psi_-$ are convergent. However, in this case the function norm is divergent, so we conclude that such modes do not exist \cite{Konoplya:2011ig}.

Now, we shall briefly relate the two standard methods used for numerical search of quasinormal modes: the shooting method and Frobenius method. These two alternative methods were used in order to guarantee the validity of the obtained results. More detailed discussion of both methods can be found in \cite{Konoplya:2011qq}.

\textbf{Shooting method}. Since at the horizon we require the purely in-going wave (\ref{hor-asymptot}), it is convenient to define a new function in such a way,
\begin{equation}\label{reg-function}
y(r)=\left(1-\frac{r_+^2}{r^2}\right)^{\imo\tilde{\omega}b}\times R(r),
\end{equation}
that it becomes regular at the event horizon.

We fix the wave-function norm so that $y(r_+)=1$. Substituting (\ref{reg-function}) into (\ref{wave-like}) and expanding the wave equation near the horizon, we find that $y'(r_+)$. This gives us boundary condition at the horizon for any fixed $\omega$, which we use for the numerical integration of the equation (\ref{wave-like}). At large distance we compare the result of our numerical integration with the large-distance asymptotic series expansion (\ref{inf-asymptot}) and find the coefficients $C_\pm$ by using the fitting procedure. Then, quasinormal modes can be found by minimizing $C_+(\omega)$ or $C_-(\omega)$ depending on the sign of $\re{\omega}$. In order to check convergence of the procedure we check that obtained frequencies do not change within specified accuracy, if we increase precision of floating-point operations or the distance at which we use the fit.

\textbf{Frobenius method.} This method allows us to find QNMs by solving numerically an equation with continued fractions, which takes much less computer time for finding quasinormal modes. One can rewrite the wave-function as
\begin{equation}\label{frobenius-prefactor}
R(r)=e^{\mp \omega jr^2}r^{\alpha_\mp}\left(\frac{r^2-r_+^2}{r^2-r_-^2}\right)^{-\imo\tilde{\omega}b}\times z(r),
\end{equation}
where $z(r)$ must be regular at spatial infinity and the event horizon, once $\omega$ is the quasinormal frequency.
We choose ``-'' sign for $\re{\omega j}>0$, and ``+'' sign for $\re{\omega j}<0$.

The function $z(r)$ can be expanded into a series near the horizon
\begin{equation}\label{frobenius}
z(r)=\sum_{i=0}^{\infty}a_i\left(\frac{r^2-r_+^2}{r^2-r_-^2}\right)^i.
\end{equation}
After we substitute (\ref{frobenius-prefactor}) and (\ref{frobenius}) into equation (\ref{wave-like}), we find the five-term recurrence relation for the coefficients $a_i$. Using Gaussian eliminations we numerically reduce the five-term recurrence relation to the three-term recurrence relation \cite{Konoplya:2011qq} and solve the equation with the infinite continued fraction with respect to $\omega$ \cite{Leaver:1985ax}.

\begin{figure}
\resizebox{\linewidth}{!}{\includegraphics*{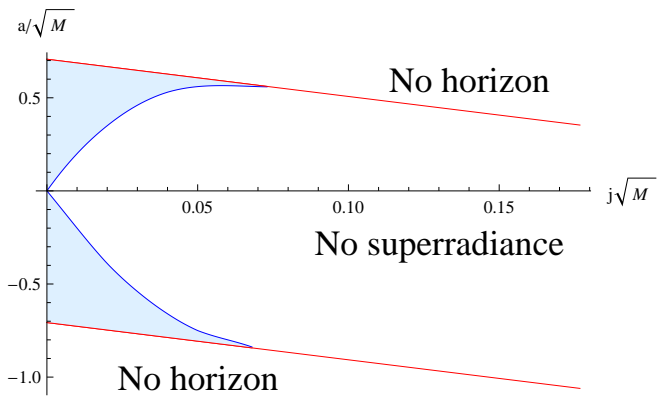}}
\caption{Region of instability (blue) for (neutral) Myers-Perry-G\"odel black holes.}\label{fig:instregion}
\end{figure}

\begin{figure}
\resizebox{\linewidth}{!}{\includegraphics*{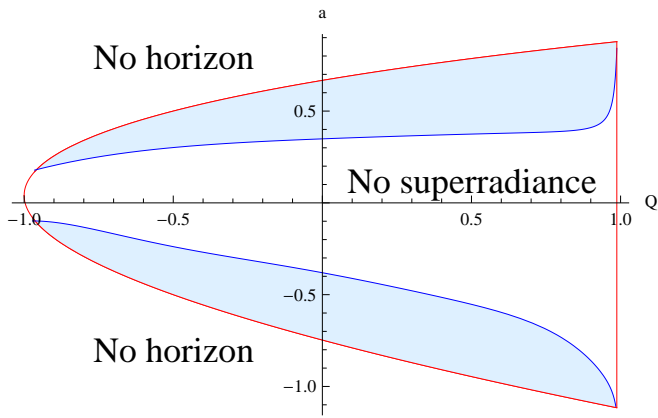}}
\caption{Region of instability (blue) for charged rotating asymptotically G\"odel black holes for $j=0.02$ (M=1).}\label{fig:instregionj002}
\end{figure}

\begin{figure}
\resizebox{\linewidth}{!}{\raisebox{-6em}{\includegraphics*{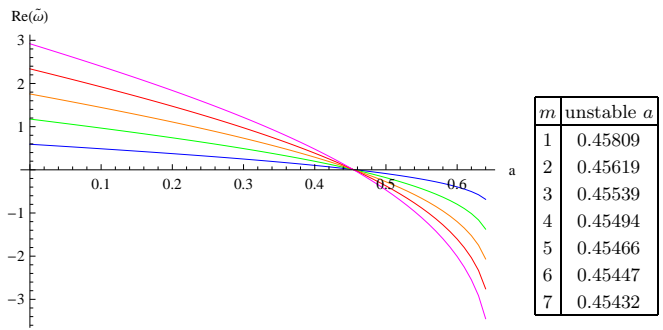}}\hspace{1em}\begin{tabular}{|c|c|}
\hline
$m$&unstable $a$\\
\hline
$1$& $0.45809$\\
$2$& $0.45619$\\
$3$& $0.45539$\\
$4$& $0.45494$\\
$5$& $0.45466$\\
$6$& $0.45447$\\
$7$& $0.45432$\\
\hline
\end{tabular}}
\caption{Real part of $\tilde{\omega}$ as functions of $a$ for $m=1$, $\lambda=2$ (blue, bottom), $m=2$, $\lambda=6$ (green), $m=3$, $\lambda=12$ (orange), $m=4$, $\lambda=20$ (red), $m=5$, $\lambda=30$ (magenta, top) ($M=1$, $Q=0$, $j=0.03$).}\label{fig:insthighm}
\end{figure}

\section{Superradiant instability}

Similarly to the asymptotically AdS rotating black holes, the superradiance occurs when
\begin{equation}\label{ineq}
\re{\omega}<m\Omega_+.
\end{equation}
On Fig. \ref{fig:instex} in the regime of small  $j$, we have shown that only the superradiant modes are unstable, while non-superradiant ones are stable.
One can also see on Fig. \ref{fig:instex} that the larger $j$ corresponds to the higher superradiant instability growth rate. The highest growth rate which we have observed is about $\sim 10^{-6}M^{-1/2}$.

For $j=0$ positive and negative $a$ produce the same quasinormal spectra (under the change $m \rightarrow -m$) because of the identical picture of ``left'' and ``right'' rotations. On Figs. \ref{fig:instregion} and \ref{fig:instregionj002} one can see that this symmetry between left and right rotations is broken due to the rotation of the universe. Now, negative $a$ allows for larger parametric region of stability than positive one. In addition, increasing of the Universe rotation ``stabilize'' the system, so that already for $j \sqrt{M} \sim 0.075 $ there is no superradiance for any $a$ (Fig. \ref{fig:instregion}).

It is well known that the Reissner-Nordstr\"om metric depends only on square of the black hole charge $Q$. Therefore, the Reissner-Nordstr\"om space-time is the same for positive and negative charges. Unlike, the RN case, the Gimone-Hashimoto metric and its rotating generalization are not ``symmetric'' respectively the change $Q \rightarrow - Q$. The dependence of the QNMs on the charge $Q$ is also not ``symmetric'' for opposite signs of the charge. Larger values of $Q$ allows for bigger parametric ``island'' of stability, so that the near extremal positively charged black hole look like having no superradiance at any
non-extremal rotation parameter $a$ (Fig. \ref{fig:instregionj002}). In addition, near extremal negative values of $Q$ evidently correspond to stability.
In the regime of large Universe scale $j$, the quasinormal frequencies approach the normal modes of the pure G\"odel space-time \cite{Konoplya:2011ig,Hiscock} similarly to the spectrum of AdS balck holes in the limit of vanishing AdS radius \cite{Konoplya:2002zu}.

As the superradiance condition (\ref{ineq}) is almost equidistant respectively $m$ for small $j$, the region of instability practically coincides for all $m$ because $\re{\omega}$ is equidistant with respect to $m$ for small $j$. Therefore, as can be seen on Fig. \ref{fig:insthighm}, when determining the threshold of instability, one can be limited by lower $m$ within acceptable accuracy.

\begin{figure*}
\resizebox{\linewidth}{!}{\includegraphics*{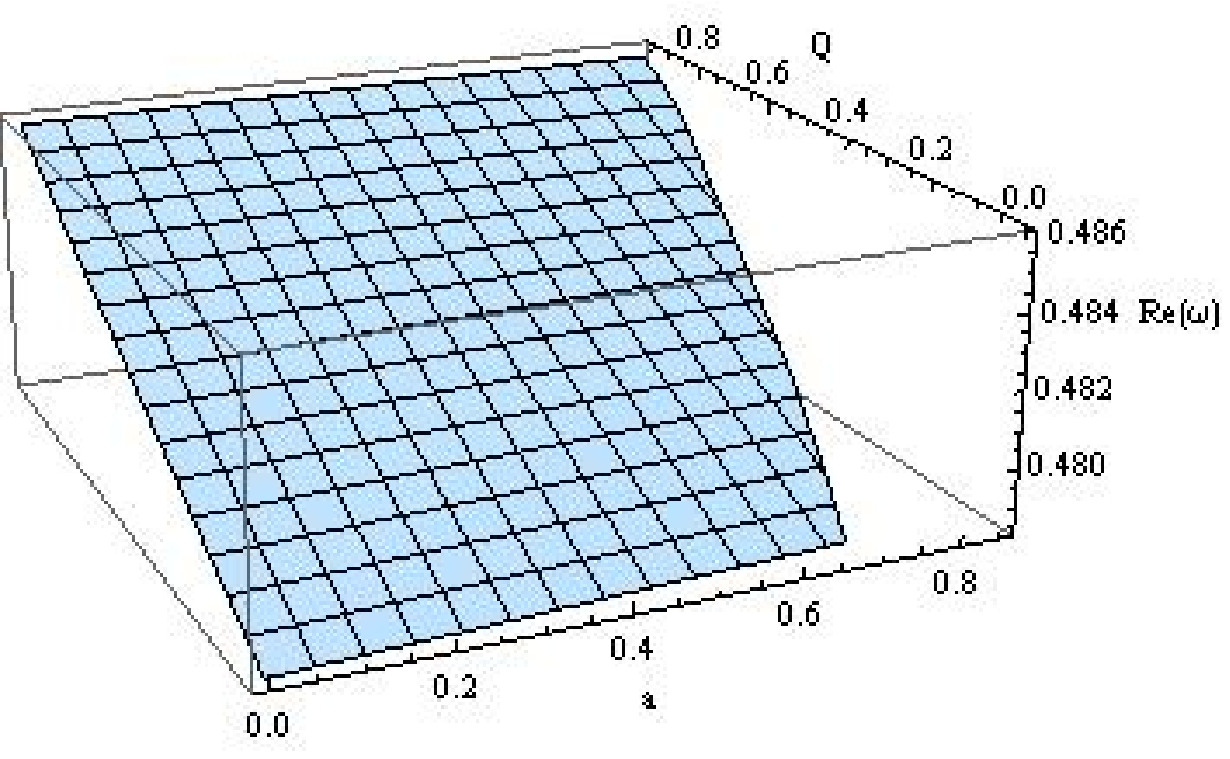}\includegraphics*{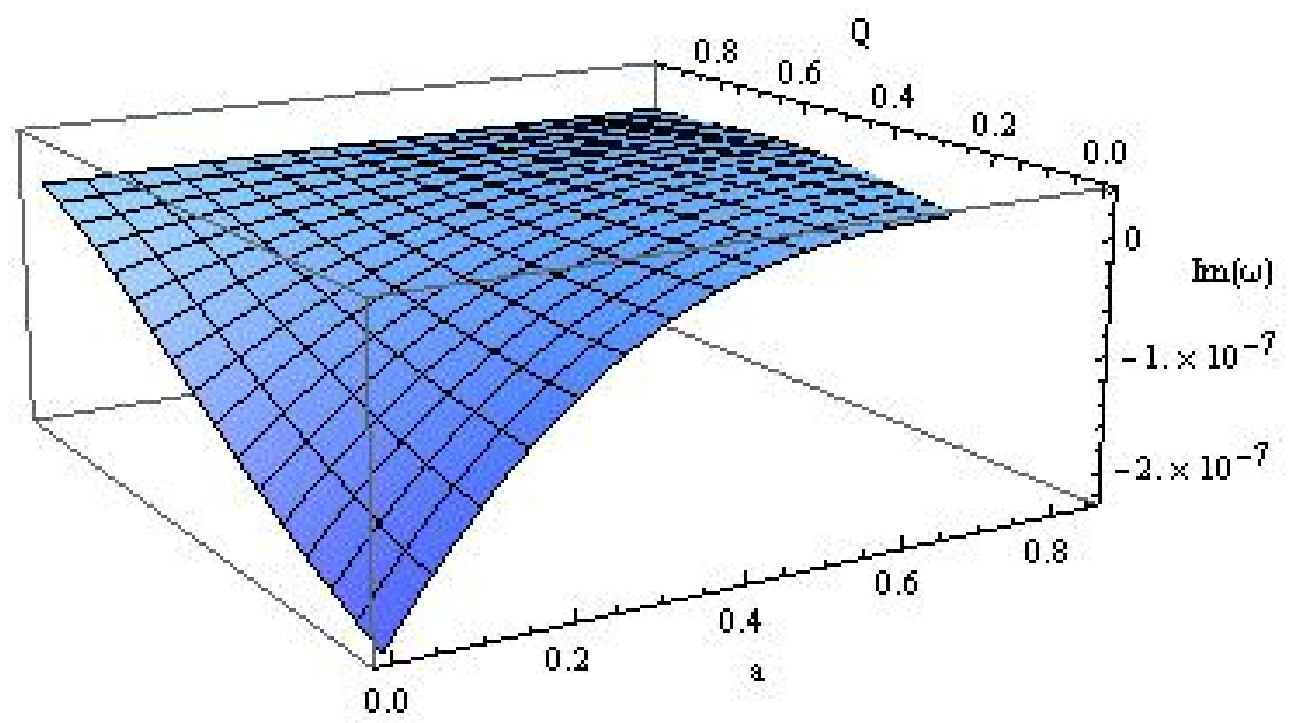}}
\caption{Real (left panel) and imaginary (right panel) parts of the dominant quasinormal mode as a function of $a$ and $Q$ for $j=0.02$ ($m=1$, $\lambda=2$, $M=1$)}\label{fig:Qaplot002}
\end{figure*}

\arXivonly{\begin{figure*}
\resizebox{\linewidth}{!}{\includegraphics*{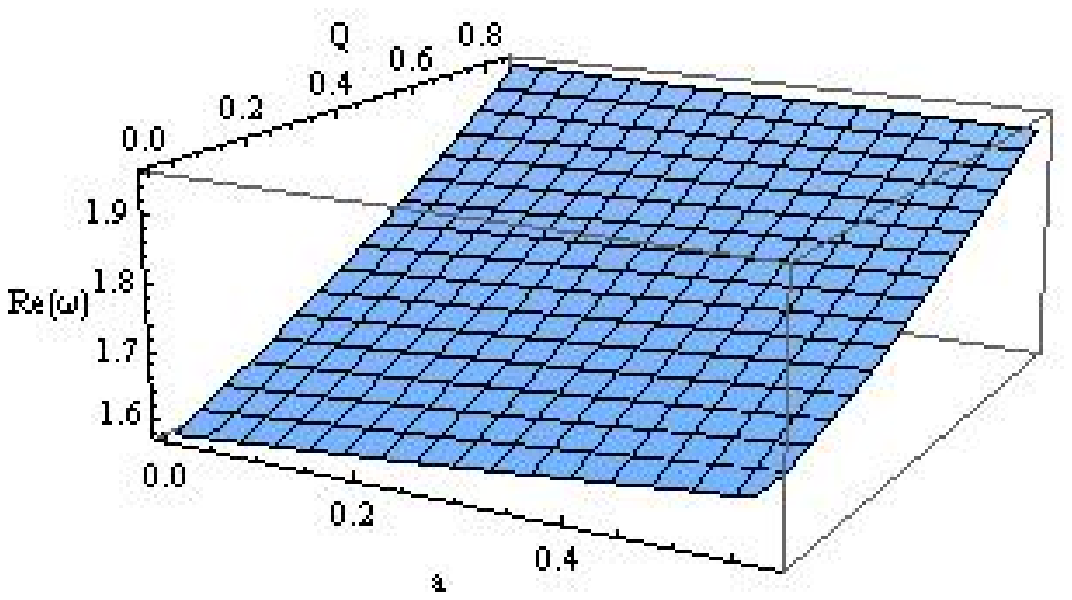}\includegraphics*{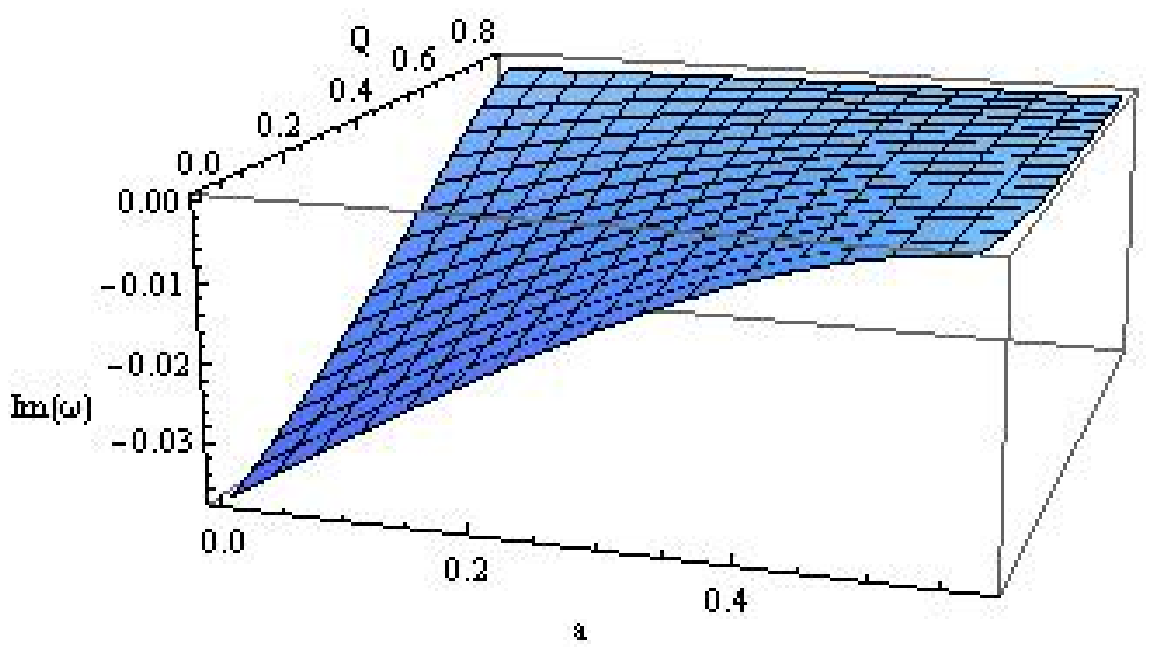}}
\caption{Real (left panel) and imaginary (right panel) parts of the dominant quasinormal mode as a function of $a$ and $Q$ for $j=0.07$ ($m=1$, $\lambda=2$, $M=1$)}\label{fig:Qaplot}
\end{figure*}

\begin{figure*}
\resizebox{\linewidth}{!}{\includegraphics*{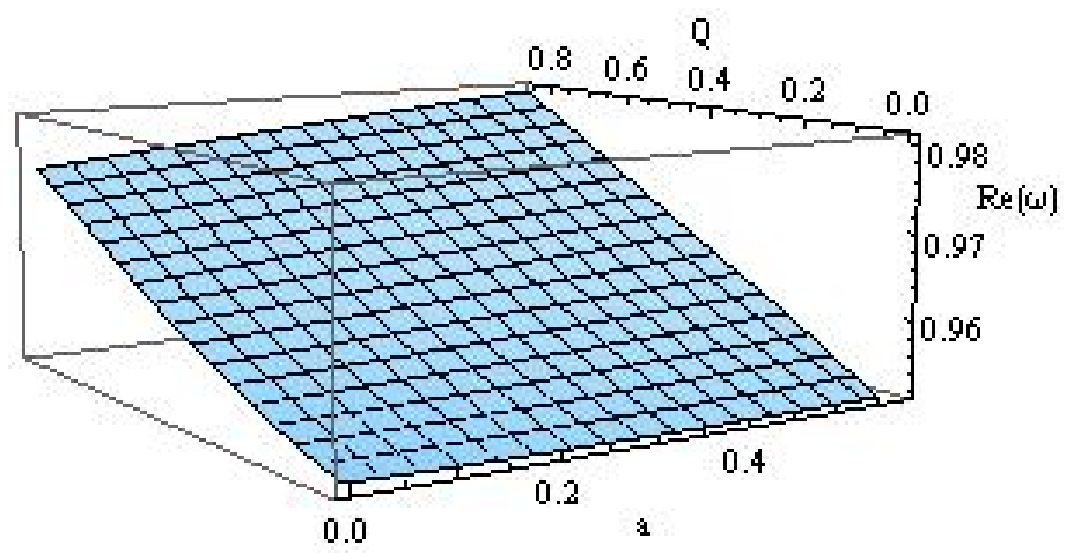}\includegraphics*{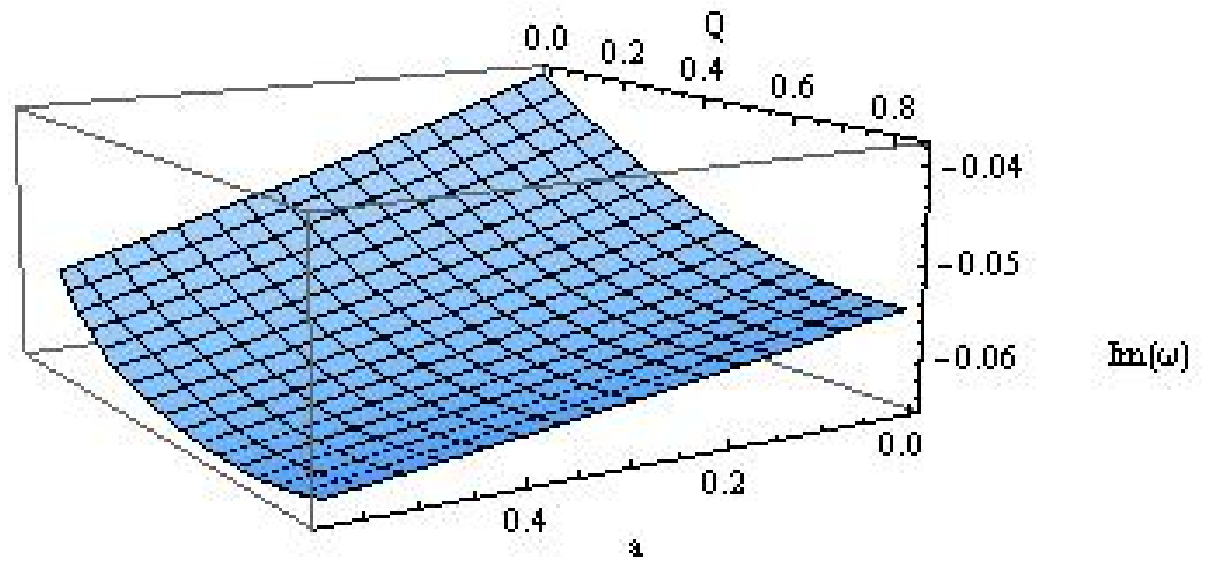}}
\caption{Real (left panel) and imaginary (right panel) parts of the dominant quasinormal mode as a function of $a$ and $Q$ for $j=0.07$ ($m=-1$, $\lambda=2$, $M=1$)}\label{fig:Qaplotm-1}
\end{figure*}

\begin{figure*}
\resizebox{\linewidth}{!}{\includegraphics*{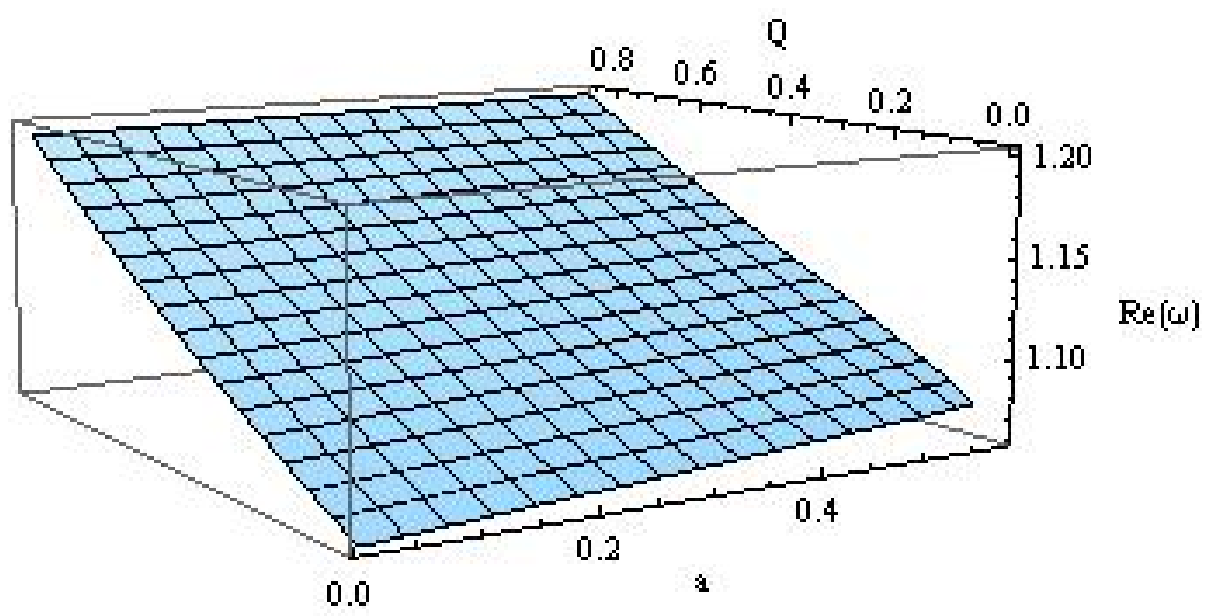}\includegraphics*{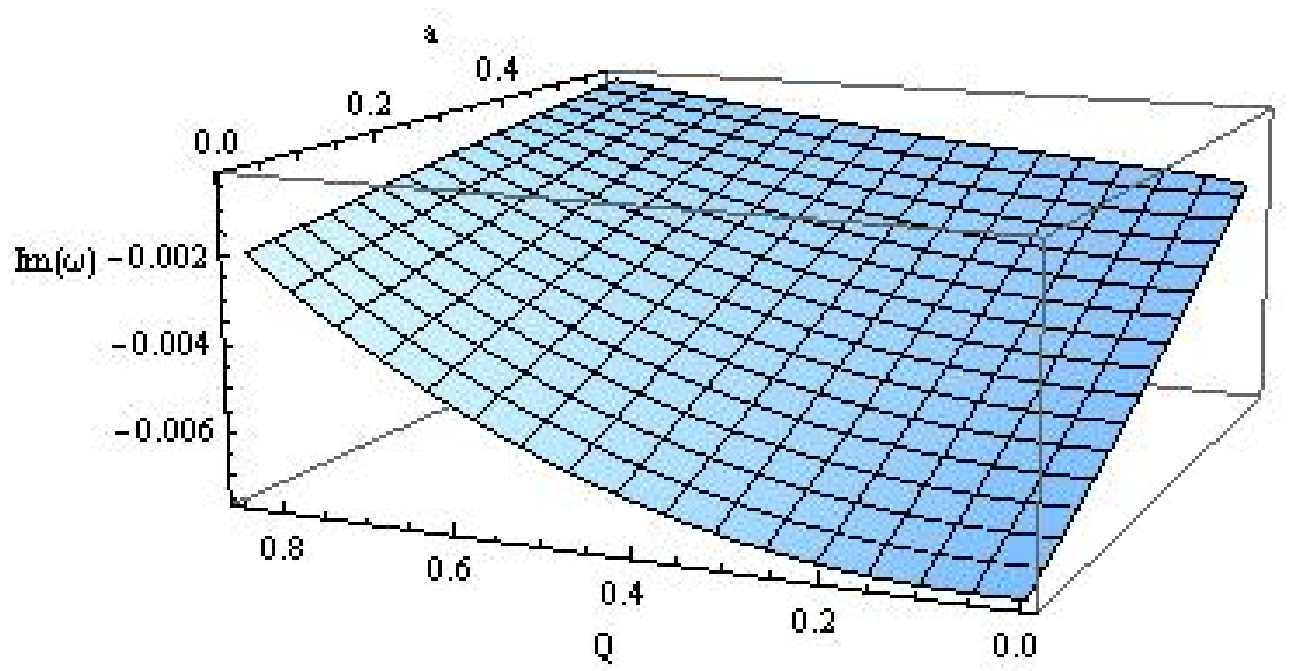}}
\caption{Real (left panel) and imaginary (right panel) parts of the dominant quasinormal mode as a function of $a$ and $Q$ for $j=0.07$ ($m=0$, $\lambda=2$, $M=1$)}\label{fig:Qaplotm0}
\end{figure*}

\begin{figure*}
\resizebox{\linewidth}{!}{\includegraphics*{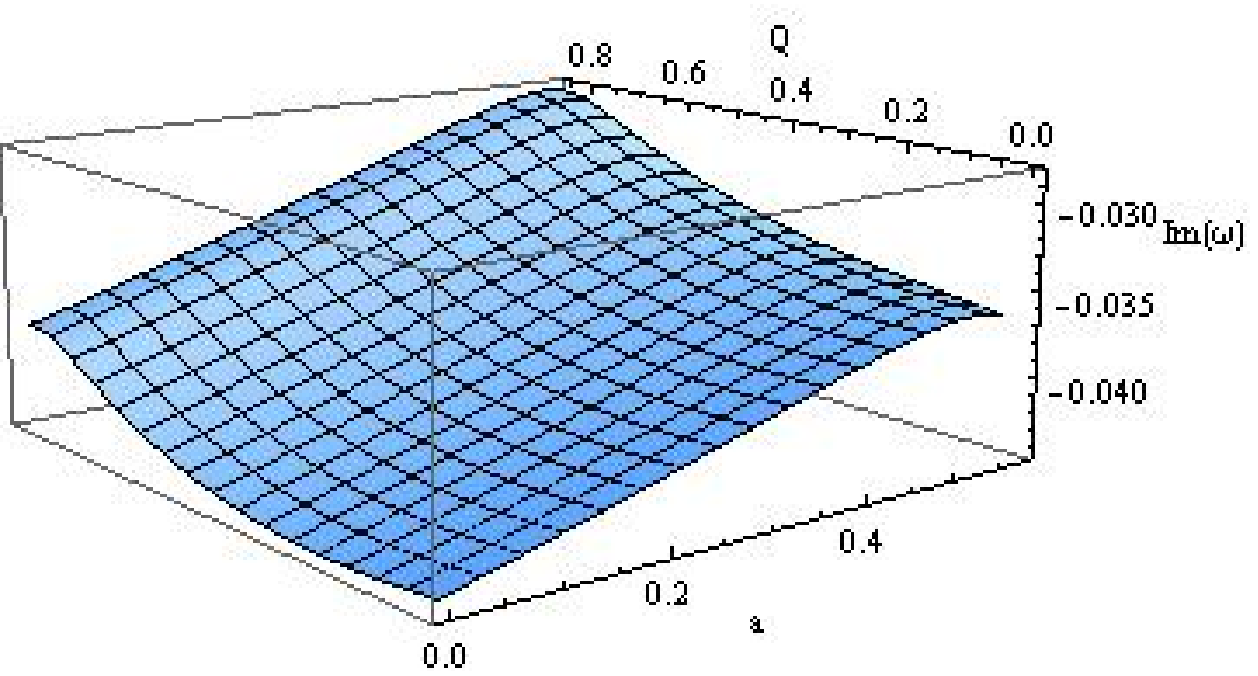}\includegraphics*{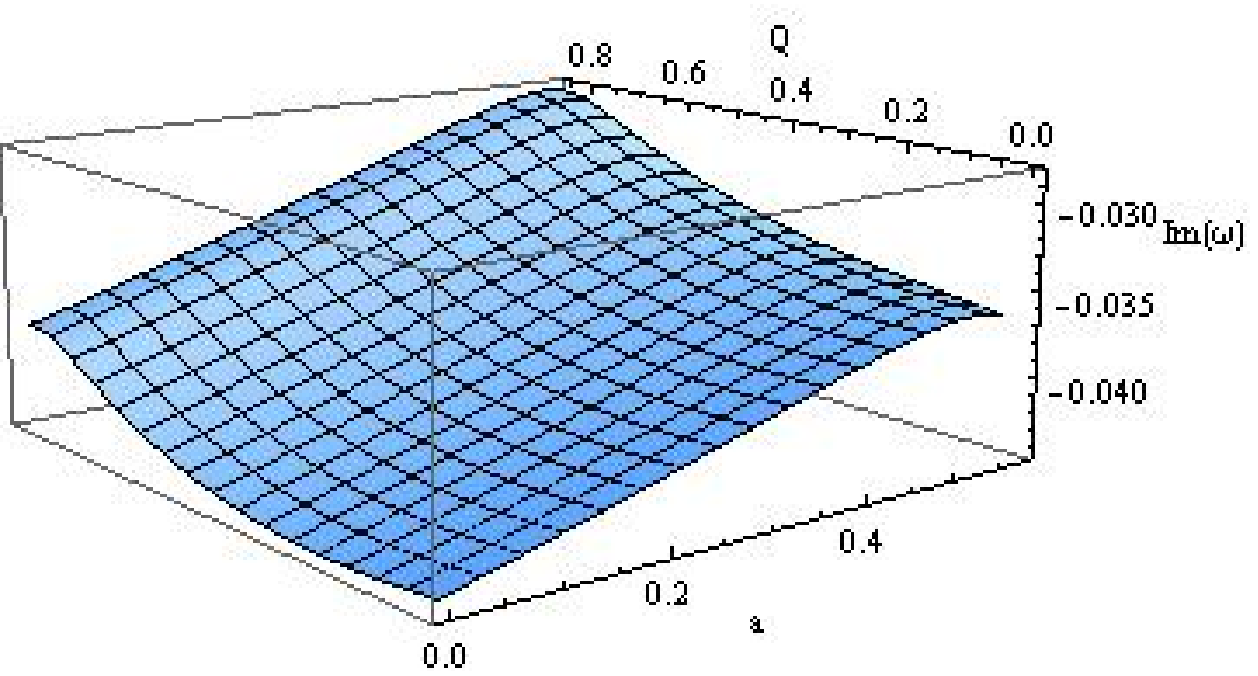}}
\caption{Real (left panel) and imaginary (right panel) parts of the dominant quasinormal mode as a function of $a$ and $Q$ for $j=0.07$ ($m=0$, $\lambda=0$, $M=1$)}\label{fig:Qaplotl0}
\end{figure*}
}%


\arXivonly{Quasinormal modes as function of the black hole charge $Q$ and the rotation parameter $a$ are shown on Figs. \ref{fig:Qaplot}, \ref{fig:Qaplotm-1}, \ref{fig:Qaplotm0}, \ref{fig:Qaplotl0}.} When increasing the black hole charge $Q$ the real oscillation frequency of the quasinormal mode $\omega$ monotonically increases (see Fig\arXivonly{s}. \ref{fig:Qaplot002}\arXivonly{, \ref{fig:Qaplot}}), while the damping rate usually decreases, making QN modes of the charged black hole
longer lived and having higher oscillation frequency. This means that charged black hole is a better oscillator (i.e. has the larger quality factor, which is proportional to $|\re{\omega}/\im{\omega}|$) than the neutral one. A similar dependence of the QNMs on the parameter of the black hole's rotation $a$ takes place, once the other parameters ($M$, $Q$, $j$) are fixed.
Therefore, in the region of stability, the charged rotating asymptotically G\"odel black hole is considerably better oscillator than the neutral non-rotating one.

\section{Conclusions}

In the present work, we have considered scalar field perturbations of the asymptotically G\"odel 5-dimensional Myers-Perry black holes with two equal angular momenta. Due to the Dirichlet boundary conditions of the wave equation which are required for the asymptotically G\"odel space-time, the quasinormal spectrum has similar features to the 5-dimensional Myers-Perry black holes in the anti-de Sitter space-time. A superradiant waves for such a rotating black hole, when immersed in an effective confining box (i.e. under Dirichlet boundary conditions) leads to the superradiant instability  for some values of the black hole parameters. A similar kind of instability we have observed for the Myers-Perry-G\"odel black holes. It is evident to us that the same kind of instability should occur also for gravitational perturbations, i.e. Myers-Perry-G\"odel black holes must be unstable. Further, we have accurately determined the parametric region of instability for various values of the Universe rotation, black hole rotation and the black hole charge. The found instability has a relatively small growth rate.

In our opinion, the instability found here for the particular black hole solution (the five-dimensional black hole in the presence of
the gauge field) is essentially generic phenomena which is induced only by the two factors:
\begin{enumerate}
\item the rotation of the black hole, which leads to superradiance and
\item the Dirichlet boundary condition in the asymptotic region of the rotating Universe.
\end{enumerate}
Therefore, we expect similar instability for any other types of rotating black holes immersed in a rotating G\"odel-like Universe.

Question which was not completely solved here is the stability of the extremely (positively) charged rotating black holes: Although numerical data indicates that the range of values of $a$ for which instability takes place shrink to zero when approaching the extremal positive charge $Q$, there is no analytical proof for this observation.

Another interesting question is the dynamical evolution of perturbation for the above case, which is quite non-trivial problem due to the existence of the closed time-like curves far from the black hole. There, the coordinate $t$ becomes spacelike and the azimuthal coordinate $\phi$ is timelike. This problem arises if one considers evolution of the perturbation in time domain. In the frequency domain we are free from this kind of problems and, therefore, we have considered here only self-consistent solutions.

\begin{acknowledgments}
At the initial stage this work was funded by the Alexander von Humboldt foundation (Germany) and by the Conicyt grant ACT-91: Southern Theoretical Physics Laboratory (STPLab). R. K. thanks the hospitality of The Centro de Estudios Cientifcos (CECS), Validivia, Chile. CECS is funded by the Chilean Government through the Centers of Excellence Base Financing Program of Conicyt.
A.~Z. was supported by Deutscher Academischer Austausch Dienst (DAAD) and Conselho Nacional de Desenvolvimento Cient\'ifico e Tecnol\'ogico (CNPq). A.~Z. acknowledges hospitality of the Eberhard-Karls University of T\"{u}bingen (Germany).
\end{acknowledgments}

\end{document}